\documentclass[superscriptaddress,secnumarabic,nobibnotes,aps,prd,showkeys,showpacs,nofootinbib,preprint]{revtex4}

\usepackage{graphicx}
\usepackage{float}
\usepackage{bm}
\usepackage{amsmath}
\usepackage{amsfonts}
\usepackage{amssymb}
\usepackage{epstopdf}
\usepackage{natbib}%
\newcommand{\bee}{\begin{equation}}
\newcommand{\eee}{\end{equation}}
\newcommand{\eaa}{\end{eqnarray}}
\newcommand{\baa}{\begin{eqnarray}}
\def\ni{\noindent}

\usepackage{color}

\begin{document}

\title{Tsallis' entropy, modified Newtonian accelerations \\ and the Tully-Fisher relation}

\author{Everton M. C. Abreu}\email{evertonabreu@ufrrj.br}
\affiliation{Grupo de F\' isica Te\'orica e Matem\'atica F\' isica, Departamento de F\'{i}sica, Universidade Federal Rural do Rio de Janeiro, 23890-971, Serop\'edica, RJ, Brazil}
\affiliation{Departamento de F\'{i}sica, Universidade Federal de Juiz de Fora, 36036-330, Juiz de Fora, MG, Brazil}
\author{Jorge Ananias Neto}\email{jorge@fisica.ufjf.br}
\affiliation{Departamento de F\'{i}sica, Universidade Federal de Juiz de Fora, 36036-330, Juiz de Fora, MG, Brazil}
\author{Albert C. R. Mendes}\email{albert@fisica.ufjf.br}
\affiliation{Departamento de F\'{i}sica, Universidade Federal de Juiz de Fora, 36036-330, Juiz de Fora, MG, Brazil}
\author{Alexander Bonilla}\email{abonilla@fisica.ufjf.br}
\affiliation{Departamento de F\'{i}sica, Universidade Federal de Juiz de Fora, 36036-330, Juiz de Fora, MG, Brazil}


\begin{abstract}
\noindent In this paper we have shown that the connection between the number of bits and the area of the holographic screen, where both were established in Verlinde's theory 
of entropic gravity, may depend on the thermostatistics theory previously chosen. Starting from the Boltzmann-Gibbs (BG) theory, we have reobtained the usual dependency of both, bits number and area. 
After that, using Tsallis' entropy concept within the entropic gravity formalism, we have derived another relation between the bits number 
and the holographic screen area. Moreover, we have used this new relation to derive three Newtonian-type accelerations in the context of Tsallis' statistics.  
Moreover, we have used this new relation to derive
three Newtonian-type acceleration in the context of Tsallis statistics which are
a modified gravitational acceleration, a modified MOND theory and a modified
Friedmann equation.
We have obtained the nonextensive version of the Tully-Fisher (TF) relation which shows a dependency of the distance of the star in contrast to the standard TF expression.  
The BG limit gives the standard TF law.
\end{abstract}
\date{\today}

\maketitle

The recent formalism engineered by E. Verlinde \cite{verlinde} (see also \cite{pad} for previous ideas about the issue) obtained the gravitational acceleration  by using the holographic principle \cite{thooft,susskind} and the well known equipartition law of energy.  Verlinde's ideas relied upon the fact that gravitation can be considered universal and independent of the details concerning the space-time microstructure.  Besides, he introduced new concepts involving holography since the holographic principle must unify matter, gravity and quantum mechanics \cite{thooft,susskind}.

The geometrical structure depicted in the model, is composed of a spherical surface, which plays the role of an holographic screen, and a particle of mass $M$ localized in its center. The holographic screen can be understood as a storage device for information. The number of bits (the basic unit of information in the holographic screen) is assumed to be proportional to the  area $A$ of the holographic screen and it is represented by
\begin{eqnarray}
\label{bits}
N = \frac{A }{\ell_P^2}\,\,,
\end{eqnarray}

\ni where $ A = 4 \pi r^2 $ and $\ell_P$ is the Planck length. In Verlinde's framework one can assume that the total number of bits for the energy on the screen is given by the equipartition law of energy
\begin{eqnarray}
\label{eq}
E = \frac{1}{2}\,N k_B T\,\,.
\end{eqnarray}

\ni It is important to understand that the usual equipartition theorem in this last equation can be obtained from the standard Boltzman-Gibbs (BG) thermostatistics.  Let us consider that the energy of the particle inside the holographic screen is equally divided by all bits in such a way that we can write the equation
\begin{eqnarray}
\label{meq}
M c^2 = \frac{1}{2}\,N k_B T\,\,.
\end{eqnarray}

\ni To calculate the gravitational acceleration, we can use both Eq. (\ref{bits}) and the Unruh temperature equation   \cite{unruh} which is given by
\begin{eqnarray}
\label{un}
k_B T = \frac{1}{2\pi}\, \frac{\hbar a}{c}\,\,.
\end{eqnarray}

\ni So, one can be  able to compute the  (absolute) gravitational acceleration formula
\begin{eqnarray}
\label{acc}
a &=&  \frac{l_P^2 c^3}{\hbar} \, \frac{ M}{R^2}\nonumber\\ 
&=& G \, \frac{ M}{R^2}\,\,.
\end{eqnarray}

\ni From Eq. (\ref{acc}) we can see that the Newton constant $G$ can be written in terms of the standard constants, $$G=\frac{\ell_P^2 c^3}{\hbar}\,\,.$$ 

In a nonextensive (NE) scenario, it is possible to derive the NE equipartition theorem which is written as \cite{pl}
\begin{eqnarray}
\label{ge}
E_q=\frac{1}{5-3q} N k_B T\,.
\end{eqnarray}

\noindent In Verlinde's formalism one can apply the NE equipartition equation, namely, Eq. (\ref{ge}).  In this way, we can obtain a modified acceleration formula   \cite{eu,abreu}
\begin{eqnarray}
\label{accm}
a = G_q \, \frac{ M}{R^2},
\end{eqnarray}

\ni where $G_q$ is an effective gravitational constant which can be given by
\begin{eqnarray}
\label{S}
G_q=\,\frac{5-3q}{2}\,G\,\,.
\end{eqnarray}

\ni From this last equation, we can note that the effective gravitational constant depends on the NE parameter $q$. For instance, when $q=1$ we have that $G_q=G$ (BG scenario) and for $q\,=\,5 / 3$ we have the curious and hypothetical result $G_q=0$, which defines the point $q=5/3$ as a singular one.

Tsallis' statistics \cite{tsallis}, which is the NE extension of BG statistical theory, defines a NE, i.e., nonadditive entropy as
\begin{eqnarray}
\label{nes}
S_q =  k_B \, \frac{1 - \sum_{i=1}^W p_i^q}{q-1}\;\;\;\;\;\;\qquad \Big(\,\sum_{i=1}^W p_i = 1\,\Big)\,\,,
\end{eqnarray}

\ni where $p_i$ is the probability of a system to exist within a microstate, $W$ is the total number of configurations (microstates) and 
$q$, known in the current literature as being the Tsallis parameter or NE  parameter, is a real parameter which measures the degree of nonextensivity. 

The definition of entropy in Tsallis statistics carries the standard properties of positivity, equiprobability, concavity and irreversibility. This approach has been successfully used in many different physical systems. For instance, we can mention the Levy-type anomalous diffusion \cite{levy}, turbulence in a pure-electron plasma \cite{turb} and gravitational systems \cite{sys,sa,eu}.
It is noteworthy to affirm that Tsallis thermostatistics formalism has the BG statistics as a particular case in the limit $ q \rightarrow 1$ where the standard additivity of entropy can be recovered.  In other words, it is mandatory to obtain BG statistics when we have $q \rightarrow 1$ in Tsallis approach equations.

In the microcanonical ensemble, where all the states have the same probability, Tsallis entropy reduces to \cite{sys}
\begin{eqnarray}
\label{micro}
S_q=k_B\, \frac{W^{1-q}-1}{1-q},
\end{eqnarray}
where in the limit $q \rightarrow 1$ we recover the usual BG entropy formula, $S=k_B\, \ln {W}$.

We will start our proposal by considering initially the BG statistics. From Eq.  (\ref{bits}) we will write a general relation between the bits number and the area of the holographic screen as
\begin{eqnarray}
\label{nq}
N=\frac{A}{Q}\,,
\end{eqnarray}

\ni where $Q$ is, at first sight, a parameter. We will consider the number of microstates $W$ given by
\begin{eqnarray}
\label{micro2}
W=c_1^N\,,
\end{eqnarray}

\ni where $c_1$ is the number of internal degrees of freedom of the holographic screen. In a BG statistics the entropy is given by $S=k_B\, \ln W$.
Using Eq. (\ref{micro2}) we have
\begin{eqnarray}
\label{bge2}
S=k_B N \ln c_1.
\end{eqnarray}

\ni Using the Beckenstein-Hawking formula that says that $S=k_B\,A/4l_p^2$, we can obtain the number of bits, $N$, as a function of $A$ to derive that  

\begin{eqnarray}
\label{nbg}
N=\frac{A}{4 l_p^2 \ln c_1}=\frac{A}{Q}\,,
\end{eqnarray}

\ni where $Q\equiv 4 l_p^2 \ln c_1$. If we make $4 \ln c_1=1$ then we recover the usual definition of the bits number, Eq. (\ref{bits}).   It is important to observe that, in a BG statistics scenario, the $Q$ parameter is a constant, i.e., it does not depend on the holographic screen area.

It is quite direct to generalize the BG entropy and, consequently, to obtain a more general relation for the number of bits, Eq. (\ref{nq}). To complete the task (and we can see a similar way described in references \cite{maj} and \cite{sag}), we will use Tsallis entropy in the microcanonical ensemble. Then, using Eq. (\ref{micro2}) and Tsallis entropy, Eq. (\ref{micro}), we have

\begin{eqnarray}
\label{ntsa}
k_B \,\frac{c_1^{N(1-q)}-1}{1-q}=k_B \frac{A}{4l_p^2}
\end{eqnarray}

\begin{eqnarray}
\label{ntsa2}
\Longrightarrow N=\frac{\ln\left[1+(1-q)\,\frac{A}{4l_p^2} \right]}{(1-q) \ln c_1}\,\,.
\end{eqnarray}

\ni We can recover the usual bits number formula, Eq. (\ref{bits}), for the limit $q \rightarrow 1$ if  we use $\ln c_1\,=\,1/4$ in Eq.(\ref{ntsa2}). A similar result for the value of $c_1$ was obtained by Padmanabhan in \cite{pad}.
Hence we can generalize  Eq. (\ref{nbg}) in the context of Tsallis statistics as

\begin{eqnarray}
\label{ntsa3}
N=\frac{\ln\left[1+(1-q)\,\frac{A}{4l_p^2} \right]}{\frac{1}{4}\,(1-q) }\,\,.
\end{eqnarray}

\ni It is important to mention that, for a given value of $A$, the number of bits $N$ in Eq. (\ref{ntsa3}) is now a function of the NE parameter $q$. 

Using the generalized bits number, Eq. (\ref{ntsa3}), 
together with Eqs. \eqref{ge} and \eqref{un} and defining $E_q\,=\,M c^2$,
we can obtain a modified absolute gravitational acceleration in Tsallis statistics
\begin{eqnarray}
\label{acc1}
a= G M \frac{\pi}{2 l_p^2} \; \frac{(5-3q) (1-q)}{\ln \left[1+\frac{A}{4l_p^2} (1-q)\right]}\nonumber\\= G M \frac{\pi}{2 l_p^2} \; \frac{(5-3q) (1-q)}{\ln \left[1+\frac{\pi R^2}{l_p^2} (1-q)\right]}\,\,,
\end{eqnarray}

\ni where we have used that $A$, the area of the holographic screen, is equal to $4\pi R^2$. We can see that in the limit $q \rightarrow 1$ we reobtain $a = G\, M/R^2$, which is just Eq. (\ref{acc}). In Fig. 1 we have plotted two-dimensionally, the normalized modified gravitational acceleration, 
Eq. (\ref{acc1}), as a function of the normalized radial distance $\bar{R}$ for only three different values of $q$.   In Fig. 2 we have the corresponding three dimensional graphic.   Namely, the set of curves obtained n Fig. 1 for the continuum group of $q$-values.

\begin{figure}[H]
	\begin{center}
		\label{acc2}
		\includegraphics[width=4.in, height=4.in]{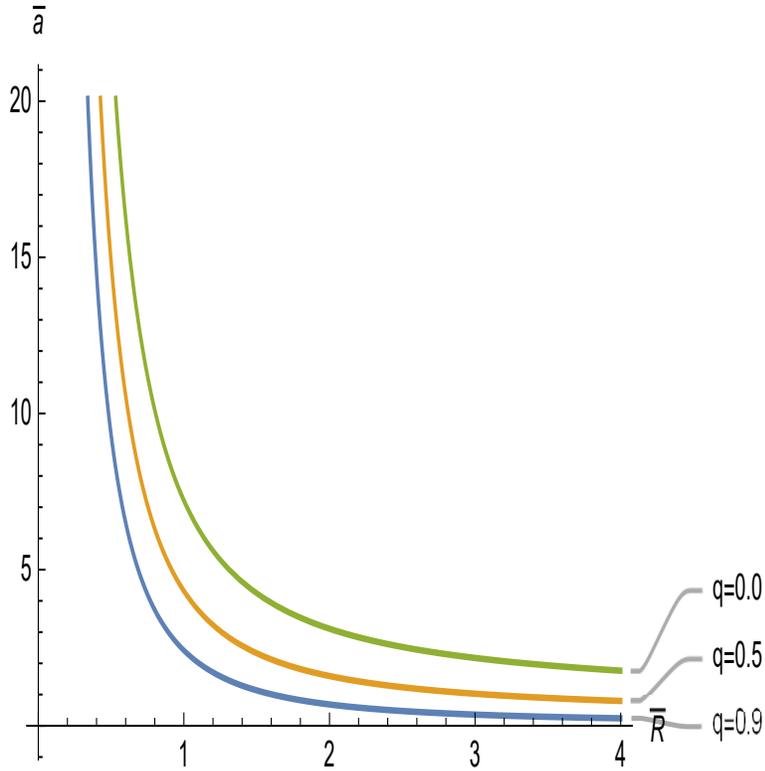}
		\caption{The modified normalized gravitational acceleration plotted as a function of the normalized radial distance $\bar{R}$ for different values of $q$. 
In the figure we have that $\bar{a}=a/(GM\pi/2l^2_P)$ and $\bar{R}=R/\sqrt{l^2_P/\pi}$.}
	\end{center}
\end{figure}

\begin{figure}[H]
	\begin{center}
		\label{acc2.1}
                \includegraphics[width=4.in, height=4.in]{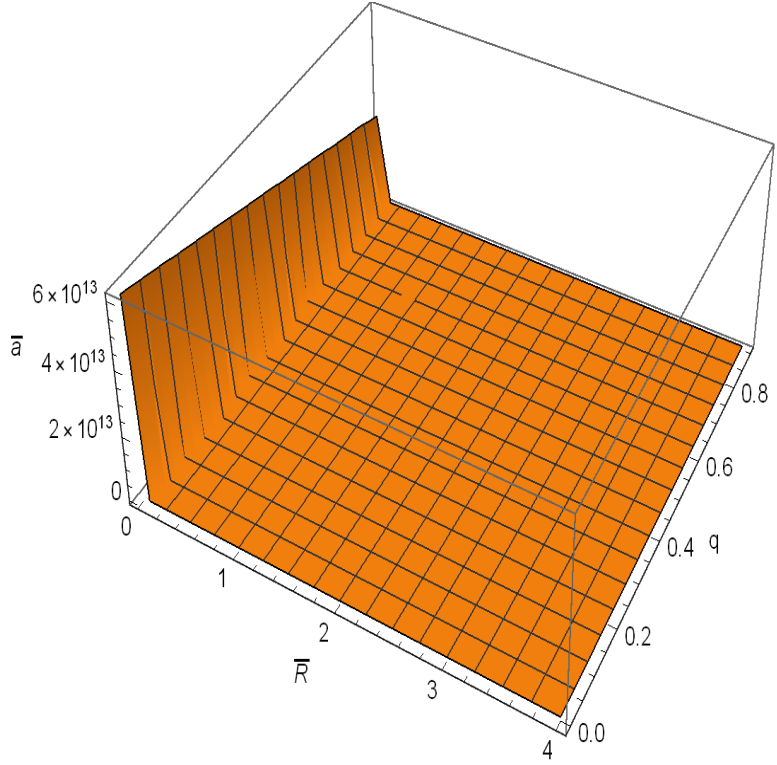}
		\caption{The modified normalized gravitational acceleration plotted in 3D  for $q$ variation from 0 to 0.9 where $\bar{R}= R/\sqrt{l_P^2/\pi}$ and $\bar{a}=a/(GM\pi/2l_P^2)$. Notice the decrease of $a$ as $q$ increases.  This figure shows the whole set (until $q=0.9$) of curves shown in figure 1. }
	\end{center}
\end{figure}

\ni From Figs. 1 and 2 we can observe clearly that the absolute value of the gravitational acceleration  decreases as the value of the NE parameter $q$ increases.  Notice that this is a new result.  Based only on Eq. (\ref{acc1}), an observer could conclude that the NE parameter is the underlying parameter that design the way massive objects attract each other. 

MOND's success \cite{mond,mond2,mond3} is due mainly to its capacity to explain the majority of the galaxies rotations.  It reproduces the well known Tully-Fisher (TF) relation \cite{tf}.  In this way, it can be an alternative to dark matter model.   However, MOND has problems to explain both the galaxy clusters temperature pattern and the confrontation with cosmology as well.  Besides, if we can say that the standard TF relation is the $q \rightarrow 1$ limit of its NE version, what is its NE version?   We will respond this question in a jiffy. 

In few words, MOND is basically a modification of Newton's second law where the force can be written as
\bee
\label{mond}
F\,=\,m\mu\Big(\frac{a}{a_0}\Big) a\,\,,
\eee

\ni where $a_0$ is a constant that will be defined just below and $\mu(x) \approx 1$ for $x>>1$ and $\mu(x) \approx x$ for $x<<1$.  For simplicity, it is usual to assume that the variation of $\mu(x)$ between the asymptotic limits happens abruptly at $x=1$ or $a=a_0$.

Concerning the rotational movement of the galaxies we have that the standard TF relation is given by 
$v^2=\sqrt{G M a_0}$, where $a_0$, the well known MOND's constant (an acceleration scale),  value is $a_0 \approx 1.2 \cdot 10^{-8}$ cm/s$^2$. 

One of us \cite{jan} has obtained MOND's theory by using Verlinde's entropic gravity approach. Basically it was considered that the fraction of bits with zero energy is given by
\begin{eqnarray}
\label{frac}
\frac{N_0}{N}=1-\frac{T}{T_c},
\end{eqnarray}

\ni where $N$ is the total bits number, $N_0$ is the number of bits with zero energy and $T_c$ is the critical temperature. For $T \geq T_c$ we have $N_0=0$.  Consequently, in MOND's regime, the number of bits with energy different from zero for a given temperature $T<T_c$ is

\begin{eqnarray}
\label{nc}
N-N_0=N \frac{T}{T_c}.
\end{eqnarray}

\ni Substituting Eq. (\ref{nc}) into (\ref{ge}) and defining $E_q=M c^2$, we have that 
\begin{eqnarray}
\label{gem}
Mc^2=\frac{1}{5-3q} N \frac{T}{T_c} k_B T.
\end{eqnarray}

\ni Then, using Eq. (\ref{bits}) and \eqref{un} into (\ref{gem}) we obtain

\begin{eqnarray}
\label{mondt}
a \left(\frac{a}{a_0}\right)=G M \frac{\pi}{2 l_p^2} \; \frac{(5-3q) (1-q)}{\ln \left[1+\frac{A}{4l_p^2} (1-q)\right]},
\end{eqnarray}

\ni and consequently we have that
\begin{eqnarray}
\label{mondt2}
a=\sqrt{G M a_0\frac{\pi}{2 l_p^2} \; \frac{(5-3q) (1-q)}{\ln \left[1+\frac{\pi R^2}{l_p^2} (1-q)\right]}},
\end{eqnarray}

\ni where $a_0$ is given by $a_0=2\pi c k_B T_c/\hbar$ and we have used $A=4\pi R^2$ in Eq. \eqref{mondt}. Therefore, we have obtained an acceleration's formula for MOND's theory in the context of Tsallis' statistics in Eq. (\ref{mondt2}).   Let us explain the physics behind this last result with more details.

Since we have the BG limit given by
\bee
\label{aa}
\lim_{q\to 1} \frac{(5-3q)(1-q)}{\ln \Big[1+(1-q)\frac{\pi R^2}{l^2_P}\Big]}\,=\,\frac{2 l^2_P}{\pi R^2}\,\,,
\eee

\ni and substituting this result into Eq. \eqref{mondt2} we have that
\bee
\label{26}
\lim_{q\to 1} a \,=\, \frac{\sqrt{GMa_0}}{R}\,\,,
\eee

\ni which is the TF relation divided by $R$.   Hence
\bee
\label{27}
\lim_{q\to 1} a \,=\, \frac{v^2}{R}\,\,,
\eee

\ni which is an expected result since it confirms MOND's primary assumption.   So, we can conclude from this last result that the acceleration given in Eq. \eqref{mondt2} is the galaxy rotation acceleration which, in the BG limit, is the galaxy centripetal acceleration.
  So, let us write that
\baa
\label{bb}
\frac{v^2}{R}\,=\,\sqrt{GMa_0 \frac{\pi}{2l^2_P} \frac{(5-3q)(1-q)}{\ln \Big[1+(1-q)\frac{\pi R^2}{l^2_P}\Big]}} \nonumber \\
\mbox{} \\
\Longrightarrow v^2\,=\, \sqrt{GMa_0 \frac{\pi R^2}{2l^2_P} \frac{(5-3q)(1-q)}{\ln \Big[1+(1-q)\frac{\pi R^2}{l^2_P}\Big]}}\,\,,
\nonumber
\eaa

\ni which can be understood as the NE version of the TF relation.  In the BG limit, 
using Eq. \eqref{aa} we have that $v^2 \,=\,\sqrt{GMa_0}$.  Notice that in Eq. \eqref{bb}, the NE TF relation is dependent of $R$.  This $R$-dependence is not a feature given by MOND, but Eq. \eqref{bb}, its NE version, is $R$-dependent.  This is a new result in the literature.  We believe that we can use Eq. \eqref{bb} to obtain valid intervals for the $q$-value.  It is an ongoing research. 

Relatively recent astronomical observations have demonstrated that, for disk galaxies, the fourth power of
the stars circular speed moving around the galaxy core, $v^4_{core}$, is proportional
to the galaxy total luminosity $L$ having an accuracy of more than
two orders of magnitude in $L$.   In other words, $v^4_{core} \propto L$. Since $L$ is proportional to
the galaxy mass, $M$, one can obtain $v^4_{core} \propto M$, i.e., the TF
law. However, we have dependency on the distance from the star until the center of
the galaxy as the Newtonian law $v^2_{core} = GM/r$ requires concerning circular motion.
In order to correct this deviation from Newton's law, astronomers suppose 
that there are halos around the galaxy, which are stuffed with dark matter
and distributed in such ways so that they satisfy the TF law for each
specific situation.  We can see clearly a different scenario from the NE version of the TF relation in Eq. \eqref{bb} where the rotation velocity depends on the $R$ distance.  
To investigate the dark matter consequences that Eq. \eqref{bb} can bring is another ongoing research that will be published elsewhere.

\begin{figure}[H]
	\begin{center}
		\label{fig:acc2.2}
		\includegraphics[width=4.in, height=4.in]{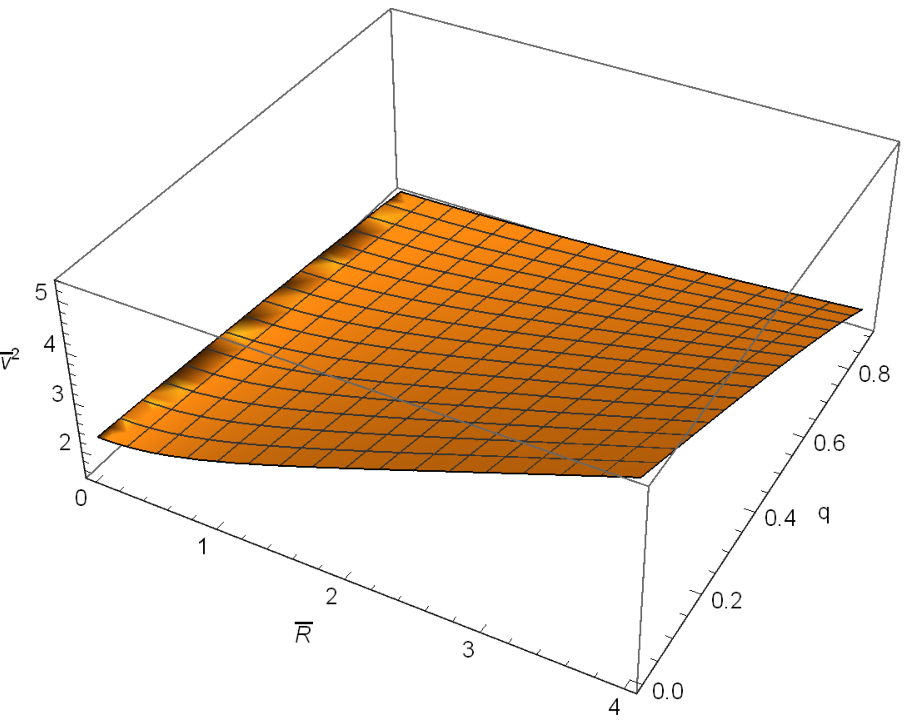}
		\includegraphics[width=4.in, height=4.in]{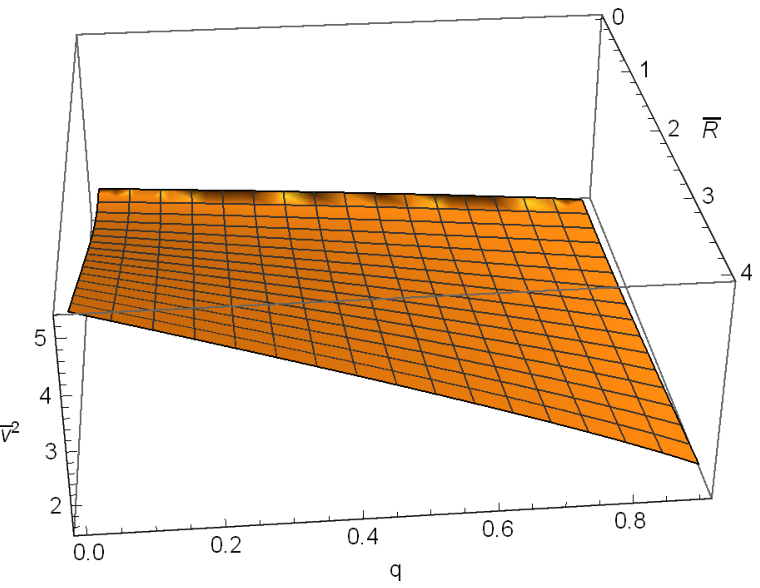}
		\caption{The nonextensive Tully-Fisher relation, i.e., the rotation velocity $\bar{v}^2 = v^2/\sqrt{GMa_0 /2}$ plotted as a function of the radial distance $\bar{R}=R/\sqrt{l_P^2/\pi}$ for different values of $q$.   For clarity's sake, both figures show different angles of the $v^2$-behavior as a function of both $q$ and $\bar{R}$.  
}
	\end{center}
\end{figure}

In Fig. $3$ we can see clearly that the NE version of the TF relation generates an $R$ dependency that does not exist in the standard TF law.  We can notice that the rotation velocity ($v^2$) increases as $R$ increase and $q$ has low values.  Near the BG limit, the velocity decreases as the $R$ also  increases together with $q$.

From Eq. (\ref{acc1}) we have the thermodynamical expression for the Newtonian acceleration. Hence,
\begin{eqnarray}
\label{accf}
\ddot{R}= \ddot{a}(t)r=G M \frac{\pi}{2 l_p^2} \; \frac{(5-3q) (1-q)}{\ln \left[1+\frac{\pi a^2 r^2}{l_p^2} (1-q)\right]}\,\,,
\end{eqnarray}

\ni where $R$ is the physical radius of the holographic screen which can be written as $R(t,r)=a(t)r$, $a(t)$ is the scale factor of the Friedmann-Robertson-Walker (FRW) metric and $r$ is the radial comoving coordinate. Based on \cite{pad2} the acceleration in Eq. (\ref{accf}) results from the active gravitational mass, which is the well-known Tolman-Komar mass \cite{tol,ko} given by
\begin{equation}
\label{tk}
M=\frac{4\pi}{3} a^3 r^3 \left(\rho+\frac{3p}{c^2}\right),
\end{equation}

\ni which is proportional to the scale function. Substituting Eq. (\ref{tk}) into Eq. (\ref{accf}) we have a modified Friedmann equation in the context of Tsallis' statistics which is

\begin{eqnarray}
\label{frie}
\frac{\ddot{a}}{a}=\frac{2\pi^2}{3l_p^2} a^2 r^2 G \left(\rho+\frac{3p}{c^2}\right)\frac{(5-3q) (1-q)}{\ln \left[1+\frac{\pi a^2 r^2}{l_p^2} 
(1-q)\right]}\,\,.
\end{eqnarray}

\ni Carrying out the limit $q\rightarrow 1$ in Eq. (\ref{frie}) we obtain the usual Friedmann equation, as expected, which is

\begin{eqnarray}
\frac{\ddot{a}}{a}=\frac{4\pi G}{3} \left(\rho+\frac{3p}{c^2}\right).
\end{eqnarray}

\ni The radial comoving coordinate $r$ in Eq. (\ref{frie}) can be eliminated using the dynamical apparent horizon radius \cite{koma} defined as
\begin{eqnarray}
\label{dr}
r=\frac{c}{H}.
\end{eqnarray}

\ni Then, using Eq. (\ref{dr}) into (\ref{frie}) and the fact that $\ddot{a}/a$ can be written as $\,\ddot{a}/a=\dot{H}+H^2$, we obtain
\begin{eqnarray}
\label{fh}
H^2 (\dot{H}+H^2)=\frac{2\pi^2}{3l_p^2} \, G \left(\rho+\frac{3p}{c^2}\right)\frac{(5-3q) (1-q)}{\ln \left[1+\frac{\pi}{l_p^2 H^2} (1-q)\right]}\,\,,
\end{eqnarray}

\ni which show us an extremely nonlinear differential equation equation for the Hubble parameter as a function of $q$.  

If we expand the denominator in the right hand side of Eq. \eqref{fh} and eliminate terms of $l^n_P$ where $n>2$, we have that
\bee
\label{fh.1}
\dot{H}+H^2\,=\, \frac{2\pi}{3} G \Big( \rho\,+\, \frac{3p}{c^2} \Big)\,(5-3q)\,\,,
\eee

\ni which is a simpler equation than \eqref{fh} and we have an homogeneous equation when $q=5/3$, our singular point from Eq. (8).   But notice that now we have $G$ in Eq. \eqref{fh.1} and not $G_q$, such as in Eq. (8).  We can rewrite Eq. \eqref{fh.1} as
\bee
\label{fh.2}
\dot{H}+H^2\,=\, \frac{4\pi}{3} G_q \Big( \rho\,+\, \frac{3p}{c^2} \Big)\,\,,
\eee

\ni where $G_q$ is given by Eq. (8).

To conclude, by choosing that the microstates number defined by $W=c_1^N$, where $c_1$ is the number of internal degrees of freedom and $N$ is the number of bits, 
we have derived a nontrivial relation for the number of bits as a function of the holographic screen area. We have used an important nongaussian statistics defined by  Tsallis' entropy. 

If we consider that the entropy of the holographic screen has a black hole form, i.e., $S=k_B A/4l_p^2$, this new number of bits leads to an interesting
modification of Newtonian acceleration, 
MOND theory and Friedmann equation. Our paper clearly shows that
when the number of bits is directly proportional to the area of the holographic screen then BG statistics is behind the previously chosen theory.

The NE version of the galaxies rotation acceleration was also obtained.   After that we have obtained the NE version of the TF relation, which relates the rotation velocity of the galaxies to its mass.  Although in the standard TF relation there is not a dependency on the distance of the star to the center of the galaxy, its NE version shows such distance dependency.  The BG limit applied in this NE TF relation has recovered the standard TF relation.  One can ask if this NE format, i.e., the $q$-dependent format, can bring any light to the dark matter problem.  This is also an ongoing research.

\section*{Acknowledgments}

\ni The authors thank CNPq (Conselho Nacional de Desenvolvimento Cient\' ifico e Tecnol\'ogico), Brazilian scientific support federal agency, for partial financial support, Grants numbers 302155/2015-5 (E.M.C.A.) and 303140/2017-8 (J.A.N.). E.M.C.A. thanks the hospitality of Theoretical Physics Department at Federal University of Rio de Janeiro (UFRJ), where part of this work was carried out.

\end{document}